\documentclass[aps,prc,groupedaddress,showpacs,manuscript]{revtex4}
\usepackage{graphicx}

\begin{document}
\title {Elliptic flow and system size dependence of transition energies at intermediate energies}
\author {Yingxun Zhang$^{1}$}%
\email{zhyx@iris.ciae.ac.cn}
\author{Zhuxia Li$^{1,2,3)}$}%
\email{lizwux@iris.ciae.ac.cn}
\address{
 1) China Institute of Atomic Energy, P. O. Box 275 (18),
Beijing 102413, P. R. China\\
2) Center of Theoretical Nuclear Physics, National Laboratory of
Lanzhou Heavy Ion Accelerator,
 Lanzhou 730000, P. R. China\\
 3) Institute of Theoretical Physics, Chinese Academic of Science,
Beijing 100080}
\date{\today}
\begin{abstract}
The elliptic flow  for $Z\le2$ particles in heavy ion collisions
at energies from several tens to several hundreds MeV per nucleon
is investigated by means of transport model,i.e. a new version of
the Improved Quantum Molecular Dynamics model (ImQMD05). In this
model, a complete Skyrme potential energy density functional is
employed. The influence of different effective interactions and
medium corrections of nucleon-nucleon cross sections on the
elliptic flow are studied. Our results show that a soft nuclear
equation of state and incident energy dependent in-medium
nucleon-nucleon cross sections are required for describing the
excitation function of the elliptic flow at intermediate energies.
The size dependence of transition energies for the elliptic flow
at intermediate energies is also studied. The system size
dependence of transition energies fits a power of system size with
a exponent of 0.223.

\end{abstract}
\pacs{25.70.-z, 24.10.Lx, 21.65.+f} \maketitle

One of the main goal for the research area of heavy ion
collisions(HICs) at intermediate energies is to extract more
accurate information on the nuclear equation of state(EoS).
Considerable progress has been made recently in determining the
equation of state of nuclear matter from heavy-ion reaction
data\cite{Dan02, Demo90,Gutb90,Reis97,Stoc86}. A prominent role
among available observables is played by collective flow. A lot of
theoretical and experimental efforts on the study of the
collective flow in HICs have been paid
\cite{Shen98,Tsang96,SWang96,Pink99,Zheng99,Lari2000,Dani2000,
Pers2002,Alt2003,Acke2001,Luka2005,Andr05,Chen05,Klak93,BALi01,Zhou94,West93,Bast97}
The elliptic flow has proven to be one of the more fruitful probes
for extracting the EoS and the dynamics of heavy ion collisions.
The parameter of elliptic flow is quantified by the second order
Fourier coefficient
$v_{2}=\langle\cos2\phi\rangle=\langle\frac{p_{x}^{2}-p_{y}^{2}}{p_{x}^{2}+p_{y}^{2}}\rangle$
from the azimuthal distribution of detected particles at
mid-rapidity as follows
\begin{equation}
\frac{dN}{d\phi}=P_{0}(1+2v_{1}\cos\phi+2v_{2}\cos2\phi)
\end{equation}
where $\phi$ is the azimuthal angle of the emitted particle
momentum relative to the x axis. Positive values for
$\langle\cos2\phi\rangle$ reflect a preferential in-plane
emission, and negative values for $\langle\cos2\phi\rangle$
reflect a preferential out-of-plane emission. The change-of-sign
recently observed at ultra-relativistic energies has received
particular interest as it reflects the increasing pressure buildup
in the non-isotropic collision zone\cite{Adler03}. Recently, the
excitation function of elliptic flow parameters at energies from
Fermi energy to relativistic energy regime for $^{197}Au+^{197}Au$
has been measured by FOPI, INDRA, ALADIN
Collabrations\cite{Luka2005,Andr05} and the transition energy from
positive to negative elliptic flow was confirmed, which is around
100 MeV per nucleon. The elliptical flow parameters
$\langle\cos2\phi\rangle$ at energies from tens to hundreds MeV
per nucleon are determined by the complex interplay among
expansion, rotation and the shadowing of spectators. Both the mean
field and two-body collision parts play important role at this
energy region. The mean field plays dominant role at low energies
and then gradually the two body collision part becomes dominant
with energy increasing. Thus, a detailed study on the excitation
function of elliptical flow at this energy region can provide more
useful information on the nucleon-nucleon interaction related to
the equation of state of nuclear matter and the medium correction
of nucleon-nucleon cross sections. The transition energy of
elliptic flow at intermediate energies may be particulary useful
in extracting the information on the nuclear effective
interaction. While the elliptic flow at the energies higher than
the transition energy will be useful to extract the medium
correction of nucleon-nucleon cross sections because two-body
collisions play more important role on collective flow at these
energie\cite{Chen05,Zhou94,Zhang05}. The another aim of this work
is to investigate the medium correction of nucleon-nucleon cross
sections through elliptic flow in heavy ion collisions at energies
from Fermi energy to relativistic energies.

In this letter, we apply the new version of Improved Quantum
Molecular Dynamics model (ImQMD05) to study the excitation
function of elliptic flow parameters for $^{197}Au+ ^{197}Au$ at
intermediate energies, and through the comparison between
measurement and model calculations to extract the information on
the effective interaction which related to EoS and the medium
correction of nucleon-nucleon cross sections. The system size
dependence of transition energies of elliptic flow from $^{58}Ni+
^{58}Ni$ to $^{197}Au+ ^{197}Au$. will also be studied.

For the convenience of readers, we first give a brief introduction
of the ImQMD05 model. The main developments of the ImQMD model
compared with the usual IQMD model are : introducing 1) the
isospin independent and dependent surface energy terms in the
energy density functional, 2)the constraint on the single particle
occupation number, and 3) the system size dependent wave packet
width\cite{Wang2004}. With the ImQMD model, it is able to
successfully describe the yields of clusters in intermediate
energy heavy ion collisions\cite{Zh05}. In the ImQMD05 model we
introduce the full Skyrme potential energy density functional
except the spin-orbit term in the local interaction part, which
allow us to choose various Skyrme interactions which describe the
ground states of nuclei and saturated nuclear matter similarly
well but predict rather different properties away from saturated
density.

In the ImQMD05 model, the nuclear local interaction potential
energy density functional $V_{loc}(\rho(\mathbf{r}))$ reads
\begin{eqnarray}
V_{loc}=\frac{\alpha }{2}\frac{\rho ^{2}}{\rho _{0}}+\frac{\beta }{\gamma +1}%
\frac{\rho ^{\gamma +1}}{\rho _{0}^{\gamma }}+\frac{g_{sur}}{2\rho _{0}}%
(\nabla \rho
)^{2}+\frac{g_{sur,iso}}{\rho_{0}}[\nabla(\rho_{n}-\rho_{p})]^{2}\nonumber\\%
+(A\rho^{2}+B\rho^{\gamma+1}+C\rho^{8/3})\delta^{2}+g_{\rho\tau}\frac{\rho^{8/3}}{\rho_{0}^{5/3}},
\label{13}
\end{eqnarray}
where $\rho$, $\rho_{n}$, $\rho_{p}$ are the nucleon, neutron, and
proton density, $\delta=(\rho_{n}-\rho_{p})/(\rho_{n}+\rho_{p})$
is the isospin asymmetry.  The first two terms in expression (2)
are the iso-scalar bulk energy part, the third term is the isospin
independent surface energy term, the forth term is the surface
symmetry energy term and the fifth term is the bulk symmetry
energy term. The last term, called the $\rho\tau$ term, is
obtained from the $\rho\tau$ term of the Skyrme potential energy
density functional by applying the Thomas-Fermi approximation to
the kinetic energy density $\tau$ and thus the explicit momentum
dependence is lost. However, the strength of this term
$g_{\rho\tau}$ is rather small compared with other iso-scalar
terms. The coefficients in expression (2) are therefore directly
related to the standard Skyrme interaction parameters as
\begin{eqnarray}
\frac{\alpha}{2}&=&\frac{3}{8}t_{0}\rho_{0},
\frac{\beta}{\gamma+1}=\frac{1}{16}t_{3}\rho_{0}^{\gamma},\nonumber\\
\frac{g_{sur}}{2}&=&\frac{1}{64}(9t_{1}-5t_{2}-4x_{2}t_{2})\rho_{0},\\
\frac{g_{sur,iso}}{2}&=&-\frac{1}{64}(3t_{1}(2x_{1}+1)+t_{2}(2x_{2}+1))\rho_{0}.\nonumber
\end{eqnarray}
And the $A$,$B$ and $C$ in the volume symmetry energy term are
also given by the Skyrme interaction parameters,
\begin{equation}
A=-\frac{t_{0}}{4}(x_{0}+1/2),B=-\frac{t_{3}}{4}(x_{3}+1/2),C=-\frac{1}{24}(\frac{3\pi^{2}}{2})^{2/3}\Theta_{sym},
\end{equation}
where $\Theta_{sym}=3t_{1}x_{1}-t_{2}(4+5x_{2})$.  The
$g_{\rho\tau}$ is determined by
\begin{equation}
g_{\rho\tau}=\frac{3}{80}(3t_{1}+(5+4x_{2})t_{2})(\frac{3\pi}{2})^{2/3}\rho_{0}^{5/3}.
\end{equation}
The $t_{0},t_{1},t_{2},t_{3}$ and $x_{0},x_{1},x_{2},x_{3}$ in
expressions (3)-(5) are the parameters of Skyrme force. In the
calculations performed in this work, for the bulk symmetry
potential energy density, we only take $\rho^{2}$ form, which is
corresponding to the form of the linear density dependence of the
symmetry potential energy as people usually use but the symmetry
energy coefficient is calculated with the full energy density
functional given by (2). The symmetry potential energy should not
play important role in the quantity studied in this work.
 Furthermore, we introduce an explicit momentum dependent term
as the same form as that in \cite{Aich87}, which reads
\begin{equation}
U_{MD}=1.57[ln(1+5\times10^{-4}\Delta p^{2})]^{2}\rho/\rho_{0}
\end{equation}
as we find the explicit momentum dependent term is important for
elliptic flow. This term provides an effective mass
$m^{*}/m$=$(1+m/pdU_{md}/dp)$, which is about 0.75 at the Fermi
momentum and about 0.95 at relative momentum around 800
AMeV\cite{Aich87}. The calculations show that without explicit
momentum dependent term, the behavior of the calculated excitation
function of elliptic flow are not consistent with that of
experiments no matter which interaction is adopted. This finding
is in agreement with the conclusion obtained in \cite {Dani2000}.
The Coulomb interaction potential energy are also introduced. By
using present model we are able to directly test effective
interactions, specifically, the various Skyrme interactions
characterized by different 'K', and '$m^{*}/m$' of EoS by
comparing the predictions of different Skyrme interactions with
measurement of elliptic flow. In this work, SkP\cite{Dob84},
SkM$^{*}$\cite{Bar82}, SLy7\cite{Chab97}, and SIII\cite{Bei75}
interactions are chosen. The first three are with similar
incompressibility $K_{\infty}\sim$ 200-229 MeV but with different
$m^{*}/m$, the last one is with $K_{\infty}\sim$ 354 MeV. Table 1
gives the parameters in the energy density functional(2) and the
properties of saturated nuclear matter for the Skyrme interactions
employed in this work.
\begin{table}[htbp]
\caption{\label{tab:table2} parameters in the ImQMD model and the
properties of saturated nuclear matter for Skryme interactions
employed in this work }
\begin{ruledtabular}
\begin{tabular}{cccccccc}
 &$SkP$ & &$SkM^{\ast}$& &$SLy7$&
 &$SIII$\\
\hline
$\alpha(MeV)$  & -356.20 && -317.40 && -293.97 && -122.75   \\
$\beta(MeV)$ &303.03 && 248.96 && 215.03 && 55.19  \\
$\gamma$ &7/6 && 7/6 && 7/6 && 2\\
$g_{sur}(MeVfm^{2})$&19.47 && 21.82 && 22.64 && 18.26 \\
$g_{sur,iso}(MeVfm^{2})$& -11.35 && -5.47 && -2.25 && -4.94\\
$g_{\rho\tau}(MeV)$& 0.00 && 5.92 && 9.92 && 6.42 \\
$m^{*}/m$& 1.00 && 0.789 && 0.687 && 0.763 \\
$\rho_{\infty}(fm^{-3})$& 0.162 && 0.160 && 0.158 && 0.145 \\
$a_{s}(MeV)$& 30.66 && 30.68 && 32.62 && 28.78 \\
$K_{\infty}(MeV)$& 200 && 216 && 229 && 354 \\
\end{tabular}
\end{ruledtabular}

\end{table}

In the collision term, the phenomenological density dependent
in-medium nucleon-nucleon cross sections are taken, which reads
\begin{equation}
\sigma_{nn}^{*}=(1-\eta\rho/\rho_{0})\sigma_{nn}^{free},
\end{equation}
where $\sigma_{nn}^{free}$ denots the free nucleon-nucleon
scattering cross sections\cite{Cug96}, which are isospin dependent.
In the treatment of Pauli-blocking in collision part, neutrons and
protons are treated separately and two criteria are used as in
\cite{QFLi2002},
\begin{equation}
\frac{4\pi}{3}r_{ij}^{3}\cdot\frac{4\pi}{3}p_{ij}^{3}\geq\frac{h^{3}}{8}
\end{equation}
and
\begin{equation}
P_{block}=1-(1-f_{i})(1-f_{j}),
\end{equation}
where $f_{i}$ is the phase space distribution function for nucleon
$i$.

The fragments are constructed by means of the coalescence model
widely used in the QMD model calculations in which particles with
relative momenta smaller than $P_{0}$ and relative distances
smaller than $R_{0}$ are coalesced into one cluster(here
$R_{0}=3.0fm$ and $P_{0}=250MeV$/c are adopted). Fig.1 shows the
charge distribution of fragments for $^{197}Au+^{197}Au$ at
$E_{beam}=60, 150, 400AMeV$ at central collisions, respectively.
The experiment data(solid symbols) are taken from
\cite{Trau04,Reis2000}. The calculation results shown are for
forward angles as the same as the experimental data. In the
calculations, the SkP Skyrme interaction is used. One sees from
the figure that the calculation results for charge distribution of
fragments are in good agreement with experimental data. Then, we
apply the ImQMD05 model to study the excitation function of
elliptic flow parameters and try to extract the information on the
effective interactions and the medium corrections of two-body
cross sections.

Fig.2 shows the excitation function of elliptic flow parameters at
mid-rapidity ($|y/y_{mc}^{proj}|\leq0.1$) for $Z\le2$ particles
for $^{197}Au+^{197}Au$ collisions at b=5fm( the reduced impact
parameter $b/b_{max}$ equals to 0.38, and
$b_{max}=1.15(A_{P}^{1/3}+A_{T}^{1/3})$). The calculated elliptic
flow is given in the rotated reference frame as the same as the
experimental data. In the figure, solid symbols denote
experimental data\cite{Gutb90,Luka2005,Andr05} and open symbols
denote calculation results with Skyrme interactions SkP, SkM*,
SLy7, and SIII, respectively. Concerning the in-medium two-body
cross sections, the $\eta$ in expression (6) is taken to be 0.2.
The general behavior of the excitation functions of elliptical
flow parameters $v_{2}$ calculated with different Skyrme
interactions are similar, i.e. the elliptic flow evolves from a
preferential in-plane(rotational like) emission($v_{2}>0$) to
out-of-plane (squeeze out) emission($v_{2}<0$) with increase of
energies. But the detailed behavior of the results from different
Skyrme interactions are rather different. One can see from the
figure that the transition energies at which the elliptic flow
parameter($v_{2}$) changes sign from positive to negative are
divergent for different Skyrme interactions. The difference is
more than 30 MeV per nucleon among the calculation results with
Skyrme interactions SkP, SkM$^{*}$, SLy7, and SIII. The transition
from preferential in-plane emission to out-of-plane emission is
because the mean field which contributes to the formation of a
rotating compound system becomes less important and the collective
expansion process based on the nucleon-nucleon scattering starts
to be predominant. The competition between the mean field and the
nucleon-nucleon collisions should strongly depend on the effective
interaction, which leads to the divergence of the transition
energies calculated with different Skyrme interactions. Clearly,
the harder EoS provides stronger pressure which leads to a
stronger out-of plane emission and thus to have a smaller
transition energy. The transition energies calculated with SkP and
SkM$^{*}$ are in agreement with experimental data while that with
SIII and SLy7 are too small compared with experimental data. To
see the relation between the elliptic flow and the EoS, in the
inset in Fig.2 we show the pressure as a function of density
calculated from the potential energy density functional (2) for
SkP, SkM$^{*}$, SLy7, SIII interactions, respectively. One can see
that the transition energy sensitively depends on the stiffness of
EoS, which depends on both K and $m^{*}/m$. Thus, the best fit to
the transition energy of the elliptic flow at intermediate
energies provides us with the information on the stiffness of EoS.
It seems to us that one needs multi-observable in order to extract
K and $m^{*}/m$ explicitly(also see\cite{Dani2000,Andr05}).

 As energy further increases $v_{2}$ becomes negative, and it reaches maximal negative
value around 400 AMeV for SkP and SkM$^{*}$ and 250 AMeV for SLy7
and SIII.  The calculations with SIII and SLy7 provide stronger
pressure at the compression zone compared with SkP and SkM$^{*}$,
which makes calculated elliptic flow to reach the maximal negative
$v_{2}$ at lower energy for SIII and SLy7. To compare the
predictions with 4 Skyrme interactions with measurement we find
that the results with SkP and SkM$^{*}$ are in reasonable
agreement with experimental data. After reaching the maximal
negative elliptic flow, the negative $v_{2}$ value decreases
again. It implies that the spectator moves faster after the
$v_{2}$ reaches the maximal negative value\cite{Andr05}. In
\cite{Reis04}, the nuclear stopping from 90 AMeV to 1.93 AGeV was
measured and maximal nuclear stopping was observed around 400 AMeV
for $^{197}Au+^{197}Au$. It seems to us that the energy for
reaching the maximal negative elliptic flow parameter is in
coincident with the energy for reaction reaching the maximal
nuclear stopping. It is clear that if the reaction system reaches
the maximal stopping around certain energies the matter formed in
the reaction should reach minimal transparency and thus particles
are preferentially out of plane emission mostly.

Now let us investigate the influence of the medium correction of
nucleon-nucleon cross sections on elliptic flow. Fig.3 shows the
excitation functions of elliptic flow parameters calculated with
$\eta=0.2, 0.0, -0.4$ in expression (6) by which we effectively
study the medium correction of nucleon-nucleon cross sections at
different nuclear environment as well as the relative momentum of
the scattering pair. The SkP Skyrme interaction is adopted in the
calculations of Fig.3. From the figure, we see that at energies
lower than transition energy the difference between the
calculation results with $\eta=0.2$ and $\eta=0.0$ is small and
both give reasonable agreement with experimental data and the
difference increases when the bombarding energy is higher than
transition energy. As energy further increases the negative
elliptic flow calculated with $\eta=0.2$ is too weak( i.e. too
small negative elliptic flow parameter). One needs a smaller
$\eta$ or even a negative $\eta$. We find a reasonable agreement
with experimental results can be obtained for the case at incident
energy around 400 AMeV when $\eta$ is taken to be about $-0.4$,
i.e. at the energy about 400 AMeV, the in-medium two-body cross
section extracted is larger than the free cross section. In
\cite{QFLi2000,QFLi2004} it was predicted that the behavior of the
in-medium elastic nucleon-nucleon cross section at super-normal
densities as a function of the relative momentum of two colliding
nuclei is first suppression and then enhancement. It is also
predicted that the in-medium elastic nucleon-nucleon cross section
increases with temperature. If we simply consider the relative
momentum of colliding nucleon pair to be roughly equal to the
relative momentum of projectile and target, and suppose the
temperature increases obviously from several tens AMeV to several
hundreds AMeV, the information on the in-medium nucleon-nucleon
cross sections extracted from the elliptic flow is qualitatively
consistent with the prediction of \cite{QFLi2000,QFLi2004}. This
study suggests that the $\eta$ in the phenomenological expression
of in-medium nucleon-nucleon cross section (6) should depend on
the reaction energy in order to mimic the medium correction of
nucleon-nucleon cross sections at different environment. To
confirm this finding we make similar calculations for the
excitation function of nuclear stopping in Au+Au at SIS energies
and compared with measurement\cite{Reis04}. The information about
the medium correction of two-body cross section extracted from
nuclear stopping is in good agreement with that obtained from
excitation function of elliptic flow in this work. The results
concerning nuclear stopping will be given in another publications.

We notice that the calculation results are not in fully agreement
with measurement at whole energy region. It means that a more
self-consistent treatment including the in-medium cross section
and the mean field, especially, a more self-consistent explicitly
momentum dependent term is needed but it is still difficult up to
now.

We further carry out the study of the system size dependence for
the elliptic flow of $Z\leq2$ particles in $^{58}Ni+^{58}Ni$,
$^{112}Sn+^{112}Sn$ $^{197}Au+^{197}Au$. We find that the
transition energies for three systems are obviously different. We
then make a systematic investigation of the system size dependence
of the transition energies of elliptic flow at intermediate energy
regime.
 Fig.4 shows the transition energies as a function of masses of combined systems.
 All reactions calculated are of symmetric
reactions, the reduced impact parameters are chosen to be 0.38.
The SkP Skryme interaction and $\eta=0.2$ in the phenomenological
expression(6) are adopted. From the figure, one sees that the
transition energy decreases with the reaction system size
increasing. One of the important reason is because the pressure
produced by Coulomb interaction increases with the system size. We
fit this curve with the following power law,
\begin{equation}
E_{tran}=x(A_{P}+A_{T})^{-\tau}.
\end{equation}
The exponent $\tau$ is about $0.223$. Here the exponent is
substantially smaller than the exponent of the size dependence of
balance energies for directed flow. Presumably, it is because more
complex effects such as the expansion of the compressed zone and
the shadowing effect of the colder spectator matter play role in
changing the sign of elliptic flow compared with the directed
flow.

In summary, we have investigated the elliptic flow in heavy ion
collisions at energies from several tens AMeV to several hundreds
AMeV with the ImQMD05 model. By changing the Skyrme interactions
we study the influence of the EoS on the elliptic flow, especially
on the transition energy and the energy that the elliptic flow
parameter reaches the maximal negative value. We find that the SkP
and SkM$^{*}$ interactions can better describe the excitation
function of elliptic flow at intermediate energies. The medium
correction of nucleon-nucleon cross sections is also studied by
changing the parameter $\eta$ in expression (6). By fitting the
experimental excitation function of elliptic flow parameters we
obtain the behavior of in-medium two nucleon cross sections as
function of relative momentum of two colliding nucleons. Our study
suggests that medium correction(the $\eta$ value) in the
phenomenological expression of in-medium cross sections should
depend on the relative momentum of colliding pair and the medium
density and temperature of nuclear medium. The linear density
dependence of the in-medium nucleon-nucleon cross section in (6)
probably be better valid when the incident energies lower than
100MeV per nucleon for HIC with heavy nuclear systems. The system
size dependence of the transition energies is investigated, which
fits a power of system size with a exponent of $0.223$.

\begin{center}
{\bf ACKNOWLEDGMENTS}
\end{center}
 This work is supported by National Natural
Science Foundation of China under Grant Nos.10175093,
10235030,10235020 and by Major State Basic Research Development
Program under Contract No.G20000774. We would like to thank Dr. P.
Danielewicz for helpful discussions.

\newpage
\begin{description}

\item[\texttt{Fig.1}]
 The charge distributions of products
in the central collisions of reactions $^{197}Au+^{197}Au$ at
$E_{beam}$=60, 150, 400AMeV calculated with the ImQMD05 model,
respectively. The SkP Skyrme interaction is chosen. The experiment
data(solid symbols) are taken from \cite{Trau04,Reis2000}. The
calculation results shown are for products at forward angles as
the same as the experimental data.

 \item[\texttt{Fig.2}] The excitation functions of elliptic flow parameters at mid-rapidity for
$Z\leq2$ particles from mid-central collisions of
$^{197}Au+^{197}Au$ calculated with SkP, $SkM^{*}$, SLy7, SIII
Skyrme interactions,respectively. The calculated results are given
in the rotated reference frame as the same as the experimental
data. The experimental data are taken from \cite{Luka2005}. The
inset shows the pressure as a function of density calculated with
SkP, $SkM^{*}$, SLy7, SIII Skyrme interactions, respectively.

\item[\texttt{Fig.3}] The excitation functions of elliptic flow
parameters at mid-rapidity for $Z\leq2$ particles from mid-central
collisions of $^{197}Au+^{197}Au$ with $\eta=0.2, 0.0, -0.4$ in
the phenomenological expression of in-medium cross sections(6),
respectively. The SkP interaction is chosen. The experimental data
are taken from \cite{Luka2005}.

\item[\texttt{Fig.4}] The transition energies for elliptic flow at
intermediate energies as a function of combined system mass.

\end{description}

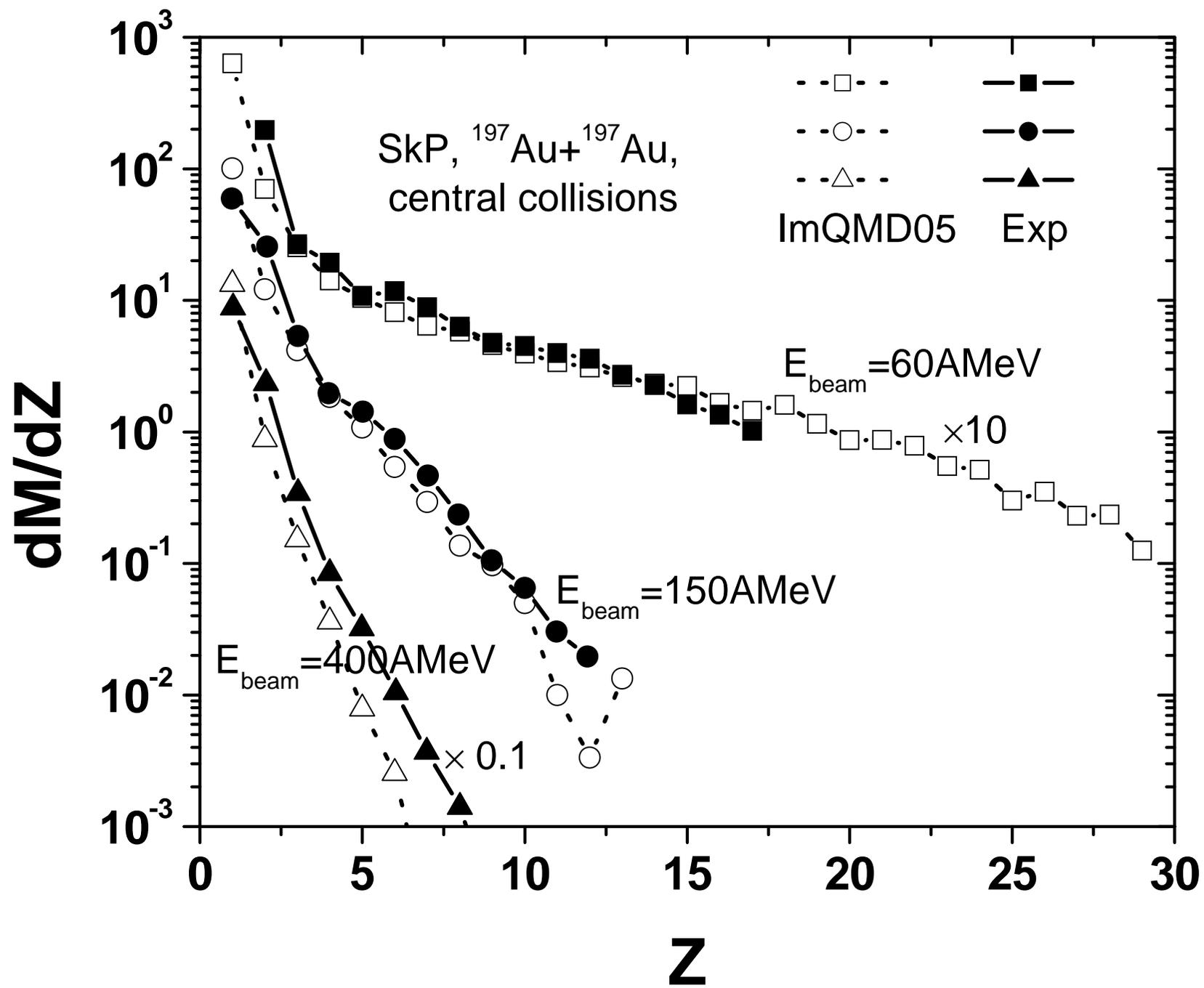
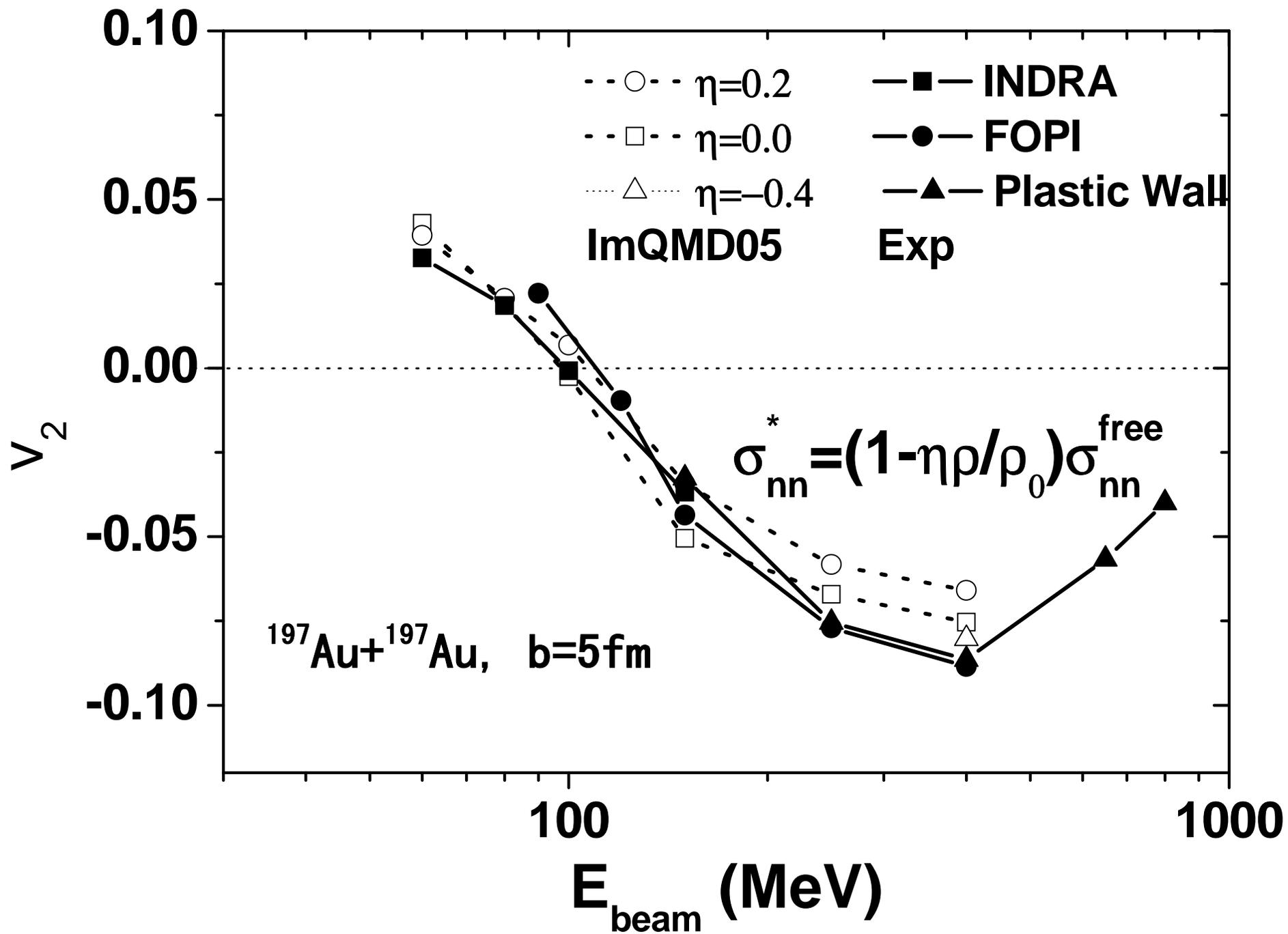
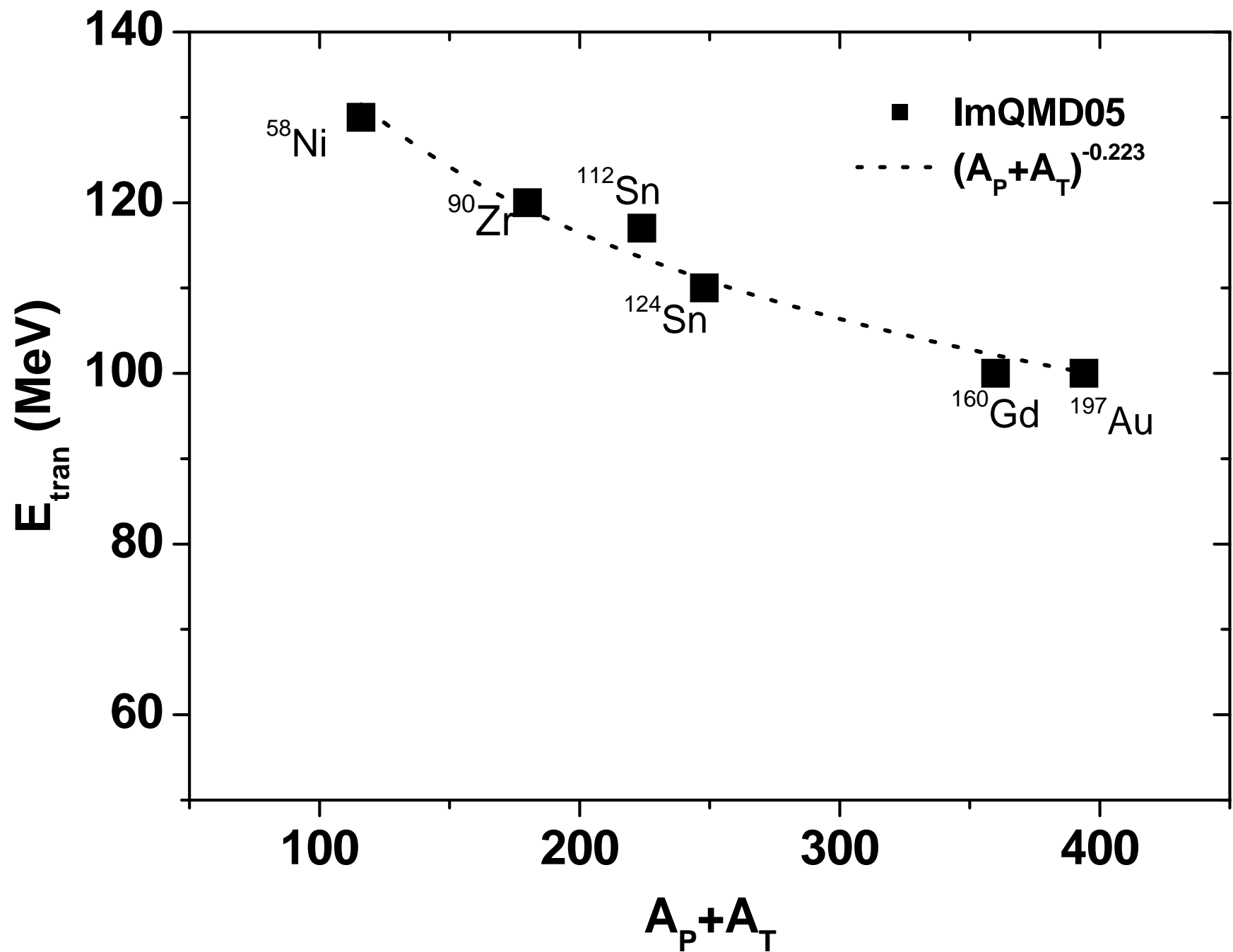
\end{document}